\documentclass[aps,twocolumn,superscriptaddress,showpacs,tightenlines]{revtex4-1}
\usepackage{epsfig,graphicx,times}
\usepackage{amstext}
\usepackage{amsmath}
\usepackage{amssymb}
\usepackage{graphicx}
\usepackage{latexsym}
\usepackage{bm}
\usepackage[colorlinks,citecolor=blue, linkcolor=blue,hyperindex,CJKbookmarks,dvipdfm]{hyperref}

\begin{document}

\title{Nonreciprocal conversion between microwave and optical photons in
electro-optomechanical systems}
\author{Xun-Wei Xu}
\email{davidxu0816@163.com}
\affiliation{Department of Applied Physics, East China Jiaotong University, Nanchang,
330013, China}
\affiliation{Beijing Computational Science Research Center, Beijing 100094, China}
\author{Yong Li}
\email{liyong@csrc.ac.cn}
\affiliation{Beijing Computational Science Research Center, Beijing 100094, China}
\affiliation{Synergetic Innovation Center of Quantum Information and Quantum Physics,
University of Science and Technology of China, Hefei, Anhui 230026, China}
\author{Ai-Xi Chen}
\email{aixichen@ecjtu.edu.cn}
\affiliation{Department of Applied Physics, East China Jiaotong University, Nanchang,
330013, China}
\author{Yu-xi Liu}
\affiliation{Institute of Microelectronics, Tsinghua University, Beijing 100084, China}
\affiliation{Tsinghua National Laboratory for Information Science and Technology
(TNList), Beijing 100084, China}
\date{\today }

\begin{abstract}
We propose to demonstrate nonreciprocal conversion between microwave
and optical photons in an electro-optomechanical system where a microwave mode
and an optical mode are coupled indirectly via two non-degenerate mechanical
modes. The nonreciprocal conversion is obtained in the broken time-reversal
symmetry regime, where the conversion of photons from one frequency to the
other is enhanced for constructive quantum interference while the conversion
in the reversal direction is suppressed due to destructive quantum
interference. It is interesting that the nonreciprocal response between the
microwave and optical modes in the electro-optomechanical system appears at
two different frequencies with opposite directions. The proposal can be used
to realize nonreciprocal conversion between photons of any two distinctive
modes with different frequencies. Moreover, the electro-optomechanical
system can also be used to construct a three-port circulator for three
optical modes with distinctively different frequencies by adding an
auxiliary optical mode coupled to one of the mechanical modes.
\end{abstract}

\maketitle


\section{Introduction}

Photons with wide range of frequencies play an important role in the quantum
information processing and quantum networks~\cite{YouJQPT05,DevoretSci13,KimbleNat08,RitterNat12}.
Microwave photons can be fast manipulated for information processing~\cite{YouJQPT05,DevoretSci13}, while the optical photons are more suitable for information transfer over long distance~\cite{KimbleNat08,RitterNat12}. However, the microwave and optical systems are not compatible with each other naturally. In order to harness the advantages of photons with different frequencies, quantum interfaces are needed to convert photons of microwave and optical modes. A hybrid quantum system should be built by combining two or more physical systems~\cite{XiangZLRMP13,JZhouSR14}.

Optomechanical (electromechnical) system is a very good candidate to serve as a quantum interface since the mechanical resonators can be easily coupled to various electromagnetic fields with distinctively different wavelengths through radiation pressure (for reviews, see Refs.~\cite{KippenbergSci08,MarquardtPhy09,AspelmeyerPT12,AspelmeyerARX13}).
In recent years, enormous progresses have been made in the optomechanical (electromechnical) systems, such as normal-mode splitting in the strong
coupling regime~\cite{GroblacherNat09,TeufelNat11a}, ground-state cooling of mechanical resonators~\cite{TeufelNat11,ChanNat11,VerhagenNat12}, and coherent state transfer between itinerant microwave (optical) fields and a mechanical oscillator~\cite{FiorePRL11,PalomakiNat13}. A hybrid electro-optomechanical system wherein a mechanical resonator is coupled to both microwave and optical modes simultaneously, provides us a quantum interface between microwave and optical systems~\cite{RegalJPCS11,TianAPB15}. It was proposed theoretically that high fidelity quantum state transfer between microwave and optical modes can be realized by using the mechanically dark mode, which is immune to mechanical dissipation~\cite{WangPRL12,TianPRL12,HKLiPRA13,ZhangKYPRL15}, and this proposal was demonstrated experimentally very soon~\cite{HillNC12,DongSci12,LiuYPRL13}. The conversion between microwave and optical fields via electro-optomechanical systems has been achieved in several different experimental setups~\cite{BochmannNPy13,BagciNat14,AndrewsNPy14} and it was shown that the wavelength conversion process is coherent and bidirectional~\cite{AndrewsNPy14}. The electro-optomechanical systems have also been studied for strong entanglement generation between microwave photon and optical photon~\cite{BarzanjehPRL12,WangYDPRL13,LTianPRL13,BarzanjehPRL15}, and such a strong continuous-variable (CV) entanglement can be exploited for the implementation of reversible CV quantum teleportation with a fidelity exceeding the no-cloning limit~\cite{BarzanjehPRL12} and microwave quantum illumination~\cite{BarzanjehPRL15}.

Nonreciprocal effect is the fundamental of isolators and circulators which are very important devices for information processing.
Such effect appears usually due to the broken time-reversal symmetry~\cite{PottonRPP04,ShadrivovNJP11}.
There are two main avenues to break the time-reversal symmetry for photons: (i) using magneto-optical
effects (e.g., Faraday rotation)~\cite{FujitaAPL00,EspinolaOL04,ZamanAPL07,HaldanePRL08,ShojiAPL08,ZWangNat09,HadadPRL10,KhanikaevPRL10,LBiNPo11,ShojiOE12} and (ii) non-magnetic strategies by employing optical nonlinearity~\cite{GalloAPL01,MingaleevJOSAB02,SoljacicOL03,RostamiOLT07,AlberucciOL08,LFanSci12,LFanOL13,AnandNL13,BiancalanaJAP08,MiroshnichenkoAPL10,CWangOE11,CWangSR12,KXiaOE13,LenferinkOE14,YYuarX14} or dynamic modulation~\cite{YuNP09,KFangNPo12,ELiNC14,DoerrOL11,DoerrOE14,LiraPRL12,KFangPRL12,TzuangNPt14,MunozPRL14,YYangOE14,WangOE10,MSKangNP11,EuterNP10,RamezaniPRA10,LFengSci11,BPengNP14,WangPRL13,JHWuPRL14,HorsleyPRL13}. Non-magnetic optical nonreciprocity based on dynamic modulation has drawn more and more attentions in recent years and many structures have been demonstrated experimentally~\cite{YuNP09,KFangNPo12,ELiNC14,DoerrOL11,DoerrOE14,LiraPRL12,TzuangNPt14,MunozPRL14,YYangOE14,MSKangNP11,EuterNP10,LFengSci11,BPengNP14,WangPRL13}
or proposed theoretically~\cite{JHWuPRL14,KFangPRL12,WangOE10,RamezaniPRA10,HorsleyPRL13}.

Nonreciprocal effect has also been developed in the context of optomechanical systems.
Optical nonreciprocal effect was proposed in an optomechanical
system consisting of an in-line Fabry-Perot cavity with one movable mirror
and one fixed mirror based on the momentum difference between forward and
backward-moving light beams~\cite{ManipatruniPRL09}. Nonreciprocity was also
studied in a microring optomechanical system when the optomechanical
coupling is enhanced in one direction and suppressed in the other one by
optically pumping the ring resonator~\cite{HafeziOE12} or by resonant
Brillouin scattering~\cite{KimNPy15,CHDongNC15}. Some of us (Xu and Li) demonstrated the possibility of optical nonreciprocal response in a three-mode optomechanical system~\cite{XuXWPRA15} where one mechanical mode is optomechanically coupled to two linearly-interacted optical modes simultaneously and the time-reversal symmetry of the system can be broken by tuning the phase difference between the two optomechanical coupling rates~\cite{KochPRA10,HabrakenNJP12,SliwaPRX15,SchmidtOpt15}. As discussed in the theoretical outlook of a recent experiment~\cite{FangKArx15}, optical nonreciprocity can be achieved in the distantly-coupled optomechanical systems with a waveguide that can mediate a tight-binding-type coupling for both the mechanical and optical cavity modes. It is worth mentioning that the two cavity modes given in Refs.~\cite{XuXWPRA15,FangKArx15} are coupled to each other directly, so that the optical modes need to be resonant or nearly resonant. On how to obtain the nonreciprocal response between two cavity modes of distinctively different wavelengths (such as a microwave mode and an optical mode), there is still a lack of studies.

More recently, Metelmann and Clerk gave a general method for generating nonreciprocal behavior in cavity-based photonic devices by employing reservoir engineering~\cite{MetelmannPRX15}.
In the spirit of the general approach of Ref.~\cite{MetelmannPRX15}, here we propose an optomechanical nonreciprocal device which allows photon routing with uni-directional links combining mechanically-mediated coherent and dissipative couplings. In our proposal, the links convert the signal carrier frequency from the microwave to the optical domain (or vice versa).
The transmission of photons from one mode to the other is determined by the quantum interference between
the two paths through the mechanically-mediated coherent and dissipative couplings. Due
to the broken time-reversal symmetry, the nonreciprocity is obtained when
the transmission of photons from one mode to the other is enhanced for
constructive quantum interference while the transmission in the reversal
direction is suppressed with destructive quantum interference. It is interesting that the electro-optomechanical system shows nonreciprocal response between the optical and microwave modes at two different
frequencies with opposite directions. Moreover, after adding an auxiliary optical mode to couple to one of the mechanical modes, the electro-optomechanical system can be used as a three-port circulator for three optical modes with distinctively different frequencies.

This paper is organized as follows: In Sec.~II, the Hamiltonian of an
electro-optomechanical system is introduced and the spectra
of the optical output fields are given. The Nonreciprocal conversion between the microwave and optical photons is shown in Sec.~III and a three-port circulator for three optical modes with distinctively different frequencies is discussed in Sec.~IV. Finally, we summarize the results in Sec.~V.

\section{Model}

\begin{figure}[tbp]
\includegraphics[bb=42 152 573 720, width=8.5 cm, clip]{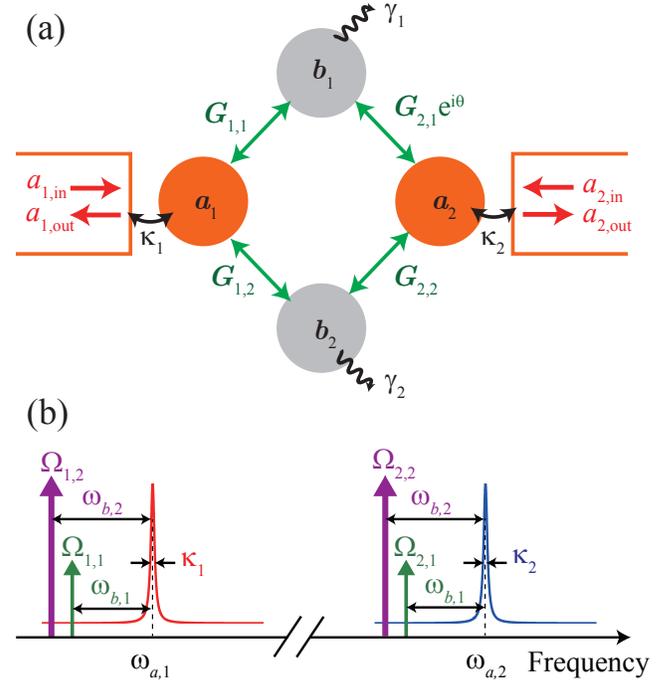}
\caption{(Color online) (a) Schematic diagram of an electro-optomechanical
system consisting of two cavity modes ($a_{1}$ and $a_{2}$) and two
mechanical modes ($b_{1}$ and $b_{2}$). The cavity mode $i$ and the mechanical mode $j$ is coupled with effective optomechanical coupling strength $G_{i,j}$ ($i,j=1,2$). (b) Schematic panel indicating the relevant frequencies involved in the nonreciprocal conversion process. The cavity mode $i$ is driven by a two-tone laser at two frequencies $\omega _{a,i}-\omega
_{b,1} $ and $\omega _{a,i}-\omega _{b,2}$ with amplitudes $\Omega _{i,1}$
and $\Omega _{i,2}$ in the well resolved sidebands ($\omega_{b,j}\gg \{\kappa _{i},\gamma _{j}\}$, where the damping rate of the mechanical mode $\gamma _{j}$ is not shown in the drawing).}
\label{fig1}
\end{figure}

As schematically shown in Fig.~\ref{fig1}(a), the electro-optomechanical system
is composed of two cavity modes (a microwave mode and an optical mode), each of which is coupled to two non-degenerate mechanical modes. The two cavity modes cannot couple to each other directly because of the vast difference of their
wavelengths. The Hamiltonian of the electro-optomechanical system is ($\hbar=1$)
\begin{eqnarray}
H_{\mathrm{eom}} &=&\sum_{i=1,2}\omega _{a,i}a_{i}^{\dag
}a_{i}+\sum_{j=1,2}\omega _{b,j}b_{j}^{\dag }b_{j}  \notag \\
&&+\sum_{i,j}g_{i,j}a_{i}^{\dag }a_{i}\left( b_{j}+b_{j}^{\dag }\right)
\notag \\
&&+\sum_{i,j}\Omega _{i,j}\left( a_{i}e^{i\left( \omega _{a,i}-\omega
_{b,j}\right) t}e^{i\phi _{i,j}}+\mathrm{H.c.}\right) ,  \label{Eq1}
\end{eqnarray}%
where $a_{i}$ ($a_{i}^{\dag }$) is the bosonic annihilation (creation)
operator of the cavity mode $i$ with resonance frequency $\omega _{a,i}$, $%
b_{j}$ ($b_{j}^{\dag }$) is the bosonic annihilation (creation) operator of
the mechanical mode $j$ with resonance frequency $\omega _{b,j}$, and $g_{i,j}$ is the electromechanical (optomechanical) coupling strength between the cavity mode $i$ and the mechanical mode $j$ ($i,j=1,2$). The cavity mode $i$
is driven by a two-tone laser at two frequencies $\omega _{a,i}-\omega
_{b,1} $ and $\omega _{a,i}-\omega _{b,2}$ with amplitudes $\Omega _{i,1}$
and $\Omega _{i,2}$ in the well resolved sidebands ($\omega_{b,j}\gg \{\kappa _{i},\gamma _{j}\}$) as schematically shown in Fig.~\ref{fig1}(b), where $\kappa _{i}$ is the decay rate of the cavity mode $i$ and $\gamma _{j} $ is the damping rate of the mechanical mode $j$.
$\phi_{i,j}$ is the phase of the driving field.
We can write each operators for the cavity modes as the
sum of its quantum fluctuation operator and classical mean value, $a_{i}\rightarrow a_{i}+\alpha_{i}(t)$. In the condition that \textrm{min}$\left[ \omega_{b,j},\left\vert \omega _{b,1}-\omega _{b,2}\right\vert \right] \gg \mathrm{%
max}\left[ \left\vert g_{i,j}\alpha _{i}(t)\right\vert \right] $, the classical part $\alpha_{i}(t)$ can be given approximately as $\alpha_{i}(t)\approx \sum_{j=1,2}\alpha _{i,j}e^{i\omega _{b,j}t}$, where the classical amplitude $\alpha _{i,j}$ is determined by solving the classical equation of motion with only cavity drive $\Omega _{i,j}$ at frequency $\omega _{a,i}-\omega _{b,j}$~\cite{WangYDPRL13,LTianPRL13,KronwaldPRA13,OjanenPRA14}. To linearize the Hamiltonian~(\ref{Eq1}), we take $\left\vert \alpha _{i,j}\right\vert \gg 1$ so that we can only keep the first-order terms in the small quantum fluctuation
operators, then the linearized Hamiltonian in the interaction picture
with respect to $H_{\mathrm{eom,}0}=\sum_{i=1,2}\omega _{a,i}a_{i}^{\dag
}a_{i}+\sum_{j=1,2}\omega _{b,j}b_{j}^{\dag }b_{j}$ is obtained as
\begin{eqnarray}
H_{\mathrm{eom,int}} &=&G_{1,1}a_{1}^{\dag }b_{1}+G_{1,1}a_{1}b_{1}^{\dag }
\notag \\
&&+G_{1,2}a_{1}^{\dag }b_{2}+G_{1,2}a_{1}b_{2}^{\dag }  \notag \\
&&+G_{2,1}e^{i\theta }a_{2}^{\dag }b_{1}+G_{2,1}e^{-i\theta
}a_{2}b_{1}^{\dag }  \notag \\
&&+G_{2,2}a_{2}^{\dag }b_{2}+G_{2,2}a_{2}b_{2}^{\dag },  \label{Eq2}
\end{eqnarray}%
where $G_{i,j}=\left\vert g_{i,j}\alpha _{i,j}\right\vert $ is the effective
electromechanical (optomechanical) coupling strength and the non-resonant and counter-rotating terms have been neglected. The phase of $\alpha_{i,j}$ can be controlled by tuning the phases $\phi_{i,j}$ of the driving fields. Actually, here the phases of $\alpha _{i,j}$ (three of them) have been absorbed by redefining the operators $a_{i}$ and $b_{j}$, and only the total phase difference $\theta $ between them has physical effects. Without loss of generality, $\theta $ is only kept in the terms of $a_{2}^{\dag }b_{1}$ and $a_{2}b_{1}^{\dag }$ in Eq.~(\ref{Eq2}) and the following derivation.

By the Heisenberg equation and taking into account the damping and corresponding noise terms, we get the quantum Langevin equations (QLEs) for the operators of the optical and mechanical modes:%
\begin{equation}
\frac{d}{dt}V\left( t\right) =-MV\left( t\right) +\sqrt{\Gamma }V_{\mathrm{in%
}}\left( t\right) ,  \label{Eq3}
\end{equation}%
with the vector $V\left( t\right) =\left(
a_{1},a_{2},b_{1},b_{2}\right) ^{T}$ of fluctuation operators, the vector $V_{\mathrm{%
in}}\left( t\right) =\left( a_{1,\mathrm{in}},a_{2,\mathrm{in}},b_{1,\mathrm{%
in}},b_{2,\mathrm{in}}\right) ^{T}$ of input operators, the diagonal damping matrix $\Gamma =%
\mathrm{diag}\left( \kappa _{1},\kappa _{2},\gamma _{1},\gamma _{2}\right) $, and the coefficient matrix
\begin{equation}
M=\left(
\begin{array}{cccc}
\frac{\kappa _{1}}{2} & 0 & iG_{1,1} & iG_{1,2} \\
0 & \frac{\kappa _{2}}{2} & iG_{2,1}e^{i\theta } & iG_{2,2} \\
iG_{1,1} & iG_{2,1}e^{-i\theta } & \frac{\gamma _{1}}{2} & 0 \\
iG_{1,2} & iG_{2,2} & 0 & \frac{\gamma _{2}}{2}%
\end{array}%
\right) .  \label{Eq4}
\end{equation}%
$a_{i,\mathrm{in}}$ and $b_{j,\mathrm{in}}$ are the input quantum fields with zero mean values.
The system is stable only if the real parts of all the eigenvalues of matrix $M$ are positive. The stability conditions can be given explicitly by using the Routh-Hurwitz criterion~\cite{DeJesusPRA87,Gradshteyn80,PaternostroNJP06,VitaliPRL07,GhobadiPRA11}. However, they are too cumbersome to be given here. All of the parameters used in the following satisfy the stability conditions.

Let us introduce the Fourier transform for an operator $o$
\begin{equation}
\widetilde{o}\left( \omega \right) =\frac{1}{\sqrt{2\pi }}\int_{-\infty
}^{+\infty }o\left( t\right) e^{i\omega t}dt,  \label{Eq5}
\end{equation}%
\begin{equation}
\widetilde{o^{\dag }}\left( \omega \right) =\frac{1}{\sqrt{2\pi }}%
\int_{-\infty }^{+\infty }o^{\dag }\left( t\right) e^{i\omega t}dt,
\label{Eq6}
\end{equation}%
then the solution to the QLEs (\ref{Eq3}) in the frequency domain can be given by
\begin{equation}
\widetilde{V}\left( \omega \right) =\left( M-i\omega I\right) ^{-1}\sqrt{%
\Gamma }\widetilde{V}_{in}\left( \omega \right),  \label{Eq7}
\end{equation}%
where $I$ denotes the identity matrix. Using the standard input-output theory~\cite{GardinerPRA85}, the Fourier transform of the output vector $V_{\mathrm{out}}\left( t\right) =\left( a_{1,\mathrm{out}},a_{2,\mathrm{out}},b_{1,\mathrm{out}},b_{2,\mathrm{out}}\right) ^{T}$ is obtained as~\cite{AgarwalPRA12}
\begin{equation}
\widetilde{V}_{\mathrm{out}}\left( \omega \right) =U\left( \omega \right)
\widetilde{V}_{\mathrm{in}}\left( \omega \right) ,  \label{Eq8}
\end{equation}%
where%
\begin{equation}
U\left( \omega \right) =\sqrt{\Gamma }\left( M-i\omega I\right) ^{-1}\sqrt{%
\Gamma }-I.  \label{Eq9}
\end{equation}%
The spectrum of the field with operator $o$ is defined as
\begin{equation}
s_{o}\left( \omega \right) =\int_{-\infty
}^{+\infty } d\omega ^{\prime }\left\langle
\widetilde{o^{\dag }}\left( \omega ^{\prime }\right)
\widetilde{o}\left( \omega \right) \right\rangle ,
\label{Eq10}
\end{equation}%
then the spectra of the input quantum fields, $s_{v_\mathrm{in}}\left( \omega
\right) $, are obtained as $\langle \widetilde{v_{\mathrm{in}}^{\dag }}%
\left( \omega ^{\prime }\right) \widetilde{v_{\mathrm{in}}}\left( \omega
\right) \rangle =s_{v_\mathrm{in}}\left( \omega \right) \delta \left(
\omega +\omega ^{\prime }\right) $ and $\langle \widetilde{v_{\mathrm{in%
}}}\left( \omega ^{\prime }\right) \widetilde{v_{\mathrm{in}}^{\dag }}\left(
\omega \right) \rangle =\left[ 1+s_{v_\mathrm{in}}\left( \omega
\right) \right] \delta \left( \omega +\omega ^{\prime }\right) $, where the
term \textquotedblleft 1\textquotedblright\ results from the effect of
vacuum noise and $\widetilde{v_{\mathrm{in}}^{\dag }}$ ($\widetilde{v_{%
\mathrm{in}}}$) is the Fourier transform of $v_{\mathrm{in}}^{\dag }$ ($v_{%
\mathrm{in}}$) (for $v_{\mathrm{in}}=a_{1,\mathrm{in}},a_{2,\mathrm{in}},b_{1,\mathrm{in}},b_{2,\mathrm{in}}$).
The relation between the vector of the spectrum of the output fields $S_{\mathrm{out}}\left( \omega \right)$ and the vector of the spectrum of the input fields $S_{\mathrm{in}}\left( \omega \right)$ is given by
\begin{equation}
S_{\mathrm{out}}\left( \omega \right) =T\left( \omega \right) S_{\mathrm{in}%
}\left( \omega \right),  \label{Eq11}
\end{equation}%
where $S_{\mathrm{in}}\left( \omega \right) =\left( s_{a_{1,\mathrm{in}}%
}\left( \omega \right) ,s_{a_{2,\mathrm{in}}}\left( \omega \right) ,s_{b_{1,\mathrm{in}}}\left( \omega \right) ,s_{b_{2,\mathrm{in}}}\left( \omega
\right) \right) ^{T}$, $S_{\mathrm{out}}\left( \omega \right) =\left(
s_{a_{1,\mathrm{out}}}\left( \omega \right) ,s_{a_{2,\mathrm{out}}}\left(
\omega \right) ,s_{b_{1,\mathrm{out}}}\left( \omega \right) ,s_{b_{2,\mathrm{out}}}\left( \omega \right) \right)^{T}$. Here $T\left( \omega \right)$ is the transmission matrix with the element $T_{v,w}\left( \omega \right)$ (for $v,w=a_{1},a_{2},b_{1},b_{2}$) denoting the scattering probability from mode $w$ to mode $v$. In the next section, we will focus on the photon scattering probability between the two cavity modes. For simplicity, we define $T_{12}\left( \omega \right) \equiv T_{a_{1},a_{2}}\left( \omega \right)
=\left\vert U_{12}\left( \omega \right) \right\vert ^{2}$ and $T_{21}\left(
\omega \right) \equiv T_{a_{2},a_{1}}\left( \omega \right) =\left\vert
U_{21}\left( \omega \right) \right\vert ^{2}$, where $U_{ij}\left( \omega
\right) $ represents the element at the $i$-th row and $j$-th column of the
matrix $U\left( \omega \right) $ given in Eq.~(\ref{Eq9}).

\section{Optical nonreciprocity}

We assume that the effective optomechanical coupling strengths $G_{i,j}$, the
decay rates $\kappa _{i}$ of the cavity modes and the damping rate $\gamma
_{j}$ of the two mechanical modes satisfy the relation
\begin{equation}
\gamma _{1}\ll G_{i,j}\sim \kappa _{1}=\kappa _{2}\equiv \kappa \ll \gamma
_{2},  \label{Eq12}
\end{equation}%
i.e., the damping of the mechanical mode $1$ is much slower than the decay of
the cavity modes and this is usually satisfied; the damping of the mechanical
mode $2$ is much faster than the decay of the cavity modes and this
condition can be realized by coupling the mechanical mode $2$ to an auxiliary
cavity mode (more details are shown in next section). Under the
assumption (\ref{Eq12}), the operators of the mechanical mode $2$ can be eliminated from QLE~(\ref{Eq3}) adiabatically~\cite{JahnePRA09,XWXuPRA15}, then we have
\begin{equation}
\frac{d}{dt}V^{\prime }\left( t\right) =-M^{\prime }V^{\prime }\left(
t\right) +\sqrt{\Gamma ^{\prime }}V_{\mathrm{in}}^{\prime }\left( t\right) -i%
\sqrt{\Lambda }b_{2,\mathrm{in}},  \label{Eq13}
\end{equation}%
with the vector $V^{\prime }\left( t\right) =\left(
a_{1},a_{2},b_{1}\right) ^{T}$ of fluctuation operators, the vector $V_{\mathrm{in}%
}^{\prime }\left( t\right) =\left( a_{1,\mathrm{in}},a_{2,\mathrm{in}},b_{1,%
\mathrm{in}}\right) ^{T}$ of input operators, the diagonal damping matrices $\Gamma ^{\prime }=%
\mathrm{diag}\left( \kappa _{1},\kappa _{2},\gamma _{1}\right) $, $\Lambda =%
\mathrm{diag}\left( \gamma _{1,2},\gamma _{2,2},0\right) $ and the
coefficient matrix
\begin{equation}
M^{\prime }=\left(
\begin{array}{ccc}
\frac{\kappa _{1}+\gamma _{1,2}}{2} & J_{2} & iG_{1,1} \\
J_{2} & \frac{\kappa _{2}+\gamma _{2,2}}{2} & iG_{2,1}e^{i\theta } \\
iG_{1,1} & iG_{2,1}e^{-i\theta } & \frac{\gamma _{1}}{2}%
\end{array}%
\right),  \label{Eq14}
\end{equation}%
where the dissipative coupling strength $J_{2}=2G_{1,2}G_{2,2}/\gamma _{2}$, and the
decay rates $\gamma _{1,2}=4G_{1,2}^{2}/\gamma _{2}$ and $\gamma_{2,2}=4G_{2,2}^{2}/\gamma _{2}$ are induced by the mechanical mode $2$. Using
the Fourier transform and the standard input-output relation, we can get the
output vector $V_{\mathrm{out}}^{\prime }\left( t\right) =\left( a_{1,%
\mathrm{out}},a_{2,\mathrm{out}},b_{1,\mathrm{out}}\right) ^{T}$ in the
frequency domain as
\begin{equation}
\widetilde{V}_{\mathrm{out}}^{\prime }\left( \omega \right) =U^{\prime
}\left( \omega \right) \widetilde{V}_{\mathrm{in}}^{\prime }\left( \omega
\right) -iL^{\prime }\left( \omega \right) b_{2,\mathrm{in}},  \label{Eq15}
\end{equation}%
where%
\begin{equation}
U^{\prime }\left( \omega \right) =\sqrt{\Gamma ^{\prime }}\left( M^{\prime
}-i\omega I\right) ^{-1}\sqrt{\Gamma ^{\prime }}-I,  \label{Eq16}
\end{equation}%
\begin{equation}
L^{\prime }\left( \omega \right) =\sqrt{\Gamma ^{\prime }}\left( M^{\prime
}-i\omega I\right) ^{-1}\sqrt{\Lambda }.  \label{Eq17}
\end{equation}

The explicit expressions of the transmission coefficients between the two
cavity modes are of the form
\begin{equation}
U_{12}^{\prime }\left( \omega \right) =\frac{-\sqrt{\kappa _{1}\kappa _{2}}%
\left( J_{1}^{\prime }+J_{2}\right) }{D\left( \omega \right) } ,
\label{Eq18}
\end{equation}%
\begin{equation}
U_{21}^{\prime }\left( \omega \right) =\frac{-\sqrt{\kappa _{1}\kappa _{2}}%
\left( J_{1}+J_{2}\right) }{D\left( \omega \right) } ,  \label{Eq19}
\end{equation}%
where
\begin{eqnarray}
D\left( \omega \right) &=&\left[ \frac{\kappa _{1,\mathrm{tot}}}{2}-i\left(
\omega -\omega _{1,1}\right) \right] \left[ \frac{\kappa _{2,\mathrm{tot}}}{2%
}-i\left( \omega -\omega _{2,1}\right) \right]  \notag \\
&&-\left( J_{1}+J_{2}\right) \left( J_{1}^{\prime }+J_{2}\right) .
\label{Eq20}
\end{eqnarray}%
Here $\kappa _{i,\mathrm{tot}}$ is the total damping rate of the cavity mode $i$ given by
\begin{equation}
\kappa _{i,\mathrm{tot}}=\kappa _{i}+\gamma _{i,1}+\gamma _{i,2}.
\label{Eq21}
\end{equation}%
The $\omega$-dependent effective coupling strength $J_{1}$ ($J_{1}^{\prime }$) (coherent coupling), the effective damping rate $\gamma _{i,1}$, and the frequency shift $\omega _{i,1}$ induced by the mechanical mode $1$, are given by
\begin{equation}
J_{1}=\frac{2G_{1,1}G_{2,1}e^{i\theta }}{\gamma _{1}-i2\omega },
\label{Eq22}
\end{equation}%
\begin{equation}
J_{1}^{\prime }=\frac{2G_{1,1}G_{2,1}e^{-i\theta }}{\gamma _{1}-i2\omega },
\label{Eq23}
\end{equation}%
\begin{equation}
\gamma _{i,1}=\frac{4G_{i,1}^{2}\gamma _{1}}{\gamma _{1}^{2}+4\omega ^{2}},
\label{Eq24}
\end{equation}%
\begin{equation}
\omega _{i,1}=\frac{4G_{i,1}^{2}\omega }{\gamma _{1}^{2}+4\omega ^{2}}.
\label{Eq25}
\end{equation}%
We would like to note that the coherent coupling strength $J_{1}$ ($J_{1}^{\prime }$) and damping rates $\gamma _{i,1}$ induced by the mechanical mode $1$ are dependent on the frequency $\omega$ of the input photons, while the dissipative coupling strength $J_{2}$ and decay rates $\gamma _{i,2}$ induced by the mechanical mode $2$ are independent on the frequency $\omega$. Moreover, there are frequency shifts $\omega _{i,1}$ induced by the mechanical mode $1$ but there are almost no frequency shifts induced by the mechanical mode $2$.

Equations (\ref{Eq18}) and (\ref{Eq19}) imply that the transmission
coefficients between the two cavity modes are determined by the quantum
interference of the two paths through the mechanically-mediated coherent and dissipative couplings [i.e., $J_{1}$ ($J_{1}^{\prime }$) and $J_{2}$]. In constructive
interference, the transmission rates will be enhanced; in contrast, the
transmission rate will be suppressed with destructive interference. The
nonreciprocity is obtained in the condition that one of the transmission
coefficients [$U_{12}^{\prime }\left( \omega \right) $ or $U_{21}^{\prime
}\left( \omega \right) $] is enhanced and the other one is suppressed.
The nonreciprocity can be intuitively understood from the schematic diagram shown in Fig.~\ref{fig1}(a). The input photons from one cavity mode to the other one undergo a Mach-Zehnder-type interference: one path is the hopping through
the mechanical mode $1$ and the other path is the hopping through the
mechanical mode $2$. The phase of the first path is determined by the driven
fields as shown in Eq.~(\ref{Eq2}). The nonreciprocal response of the
electro-optomechanical system is induced by this phase, which is gauge
invariant and is associated with the broken time-reversal symmetry for the
system~\cite{KochPRA10,HabrakenNJP12,SliwaPRX15}.

The perfect nonreciprocity is obtained as $\left\vert U_{12}^{\prime
}\left( \omega \right) \right\vert =1,U_{21}^{\prime }\left( \omega \right)
=0$ or $\left\vert U_{21}^{\prime }\left( \omega \right) \right\vert
=1,U_{12}^{\prime }\left( \omega \right) =0$. In order to satisfy $%
U_{12}^{\prime }\left( \omega \right) =0$ or $U_{21}^{\prime }\left( \omega
\right) =0$, from Eqs.~(\ref{Eq18}) and (\ref{Eq19}), we should have
\begin{equation}
J_{1}^{\prime }=-J_{2}\text{ }\mathrm{or}\text{ }J_{1}=-J_{2}.  \label{Eq26}
\end{equation}%
Under the assumption (\ref{Eq12}), i.e., $\gamma _{1}\ll G_{i,j}\ll \gamma
_{2}$, we have
\begin{equation}
\left\vert \omega \right\vert \approx\frac{G_{1,1}G_{2,1}}{G_{1,2}G_{2,2}}%
\frac{\gamma _{2}}{2},  \label{Eq27}
\end{equation}%
and
\begin{equation}
\theta =\frac{\pi}{2} \text{ } \mathrm{or} \text{ } \frac{3\pi}{2}.  \label{Eq28}
\end{equation}%
After substituting Eq.~(\ref{Eq26}) into Eqs.~(\ref{Eq18}) and (\ref{Eq19}),
we obtain the condition for $\left\vert U_{12}^{\prime }\left( \omega \right)
\right\vert =1$ or $\left\vert U_{21}^{\prime }\left( \omega \right)
\right\vert =1$ as
\begin{equation}
\frac{8J_{2}\sqrt{\kappa _{1}\kappa _{2}}}{\left[ \kappa _{1,\mathrm{tot}%
}-i2\left( \omega -\omega _{1,1}\right) \right] \left[ \kappa _{2,\mathrm{tot%
}}-i2\left( \omega -\omega _{2,1}\right) \right] }=1.  \label{Eq29}
\end{equation}%
For simplicity we choose
\begin{equation}
\omega =\omega _{1,1}=\omega _{2,1},  \label{Eq30}
\end{equation}%
then the condition in Eq.~(\ref{Eq29}) reduces to
\begin{equation}
8J_{2}\sqrt{\kappa _{1}\kappa _{2}}=\kappa _{1,\mathrm{tot}}\kappa _{2,%
\mathrm{tot}}.  \label{Eq31}
\end{equation}%
Thus with the assumption (\ref{Eq12}), the nonreciprocity is obtained as the
effective electromechanical (optomechanical) coupling strengths satisfy the
conditions (for simplicity, we choose $G_{1,1}=G_{2,1}$ and $G_{1,2}=G_{2,2}$)
\begin{eqnarray}
G_{1,1} &=&G_{2,1}=\frac{\kappa}{2},  \label{Eq32} \\
G_{1,2} &=&G_{2,2}=\frac{\sqrt{\gamma _{2}\kappa }}{2},  \label{Eq33}
\end{eqnarray}%
and the perfect nonreciprocity appears around the frequencies
\begin{equation}
\omega =\pm \frac{\kappa}{2}.  \label{Eq34}
\end{equation}

As a specific example, under the conditions given in Eqs.~(\ref{Eq12}), (\ref{Eq32}) and (%
\ref{Eq33}), by choosing $\theta =\pi /2$, the transmission coefficients at
frequency $\omega =\kappa /2$ are given by
\begin{equation}
U_{12}^{\prime }\left( \omega \right) \approx -1,\text{ }U_{21}^{\prime
}\left( \omega \right) \approx 0,  \label{Eq35}
\end{equation}%
and the transmission coefficients at frequency $\omega =-\kappa /2$ are
given by
\begin{equation}
U_{12}^{\prime }\left( \omega \right) \approx 0,\text{ }U_{21}^{\prime
}\left( \omega \right) \approx -1.  \label{Eq36}
\end{equation}%
Under the same conditions given in Eqs.~(\ref{Eq12}), (\ref{Eq32}) and (\ref{Eq33}), if we
choose $\theta =3\pi /2$, when $\omega =\kappa /2$, the transmission
coefficients are given by
\begin{equation}
U_{12}^{\prime }\left( \omega \right) \approx 0,\text{ }U_{21}^{\prime
}\left( \omega \right) \approx -1,  \label{Eq37}
\end{equation}%
and when $\omega =-\kappa /2$, the transmission coefficients are given by
\begin{equation}
U_{12}^{\prime }\left( \omega \right) \approx -1,\text{ }U_{21}^{\prime
}\left( \omega \right) \approx 0.  \label{Eq38}
\end{equation}

\begin{figure}[tbp]
\includegraphics[bb=179 359 407 742, width=5 cm, clip]{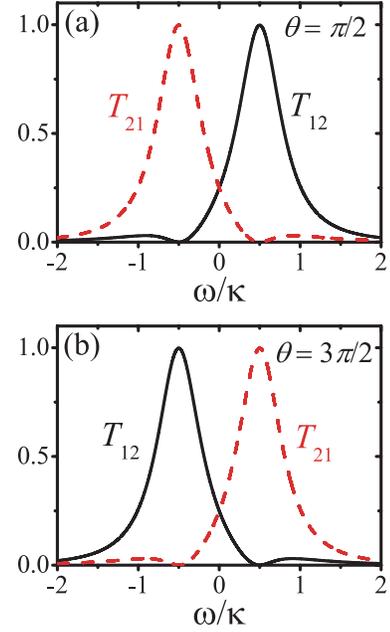}
\caption{(Color online) Scattering probabilities $T_{12}\left( \protect%
\omega \right)$ (black solid line) and $T_{21}\left( \protect\omega \right)$
(red dash line) as functions of the frequency of the incoming signal $%
\protect\omega$ for different phase difference: (a) $\protect\theta =\protect%
\pi/2$ and (b) $\protect\theta =3\protect\pi/2$. The other parameters are $%
\protect\kappa _{1}=\protect\kappa _{2}=\protect\kappa$, $\protect\gamma_{1}=%
\protect\kappa/1000$, $\protect\gamma _{2}=16\protect\kappa$, $%
G_{1,1}=G_{2,1}=\protect\kappa/2$, and $G_{1,2}=G_{2,2}=2\protect\kappa$.}
\label{fig2}
\end{figure}

In Fig.~\ref{fig2}, the scattering probabilities between the two cavity modes $%
T_{12}\left( \omega \right) =\left\vert U_{12}^{\prime }\left( \omega
\right) \right\vert ^{2}$ and $T_{21}\left( \omega \right) =\left\vert
U_{21}^{\prime }\left( \omega \right) \right\vert ^{2}$ are plotted as
functions of the frequency $\omega $ of the incoming signal for different
phase difference, where the parameters are given as $\kappa _{1}=\kappa
_{2}=\kappa $, $\gamma _{1}=\kappa /1000$, $\gamma _{2}=16\kappa $, $%
G_{1,1}=G_{2,1}=\kappa /2$, and $G_{1,2}=G_{2,2}=2\kappa $. When $\theta
\neq n\pi $ ($n$ is an integer), the time-reversal symmetry is broken and the
electro-optomechanical system exhibits a non-reciprocal response. The
optimal optical nonreciprocal response is obtained when $\theta =\pi /2$ or $\theta =3\pi /2$. As shown in Fig.~\ref{fig2}, the electro-optomechanical system shows
nonreciprocal response between the optical and microwave modes at two
different frequencies with opposite directions: when $\theta =\pi /2$ as
shown in Fig.~\ref{fig2} (a), we have $T_{21}\left( \omega \right) \approx 1$%
, $T_{12}\left( \omega \right) \approx 0$ at $\omega =-\kappa /2$ and $%
T_{12}\left( \omega \right) \approx 1 $, $T_{21}\left( \omega \right)
\approx 0$ at $\omega =\kappa /2$; when $\theta =3\pi /2$ as shown in Fig.~%
\ref{fig2} (b), we have $T_{12}\left( \omega \right) \approx 1$, $%
T_{21}\left( \omega \right) \approx 0$ at $\omega =-\kappa /2$ and $%
T_{21}\left( \omega \right) \approx 1$, $T_{12}\left( \omega \right) \approx
0$ at $\omega =\kappa /2$.

\section{Optical circulator}

\begin{figure}[tbp]
\includegraphics[bb=20 179 578 647, width=8.5 cm, clip]{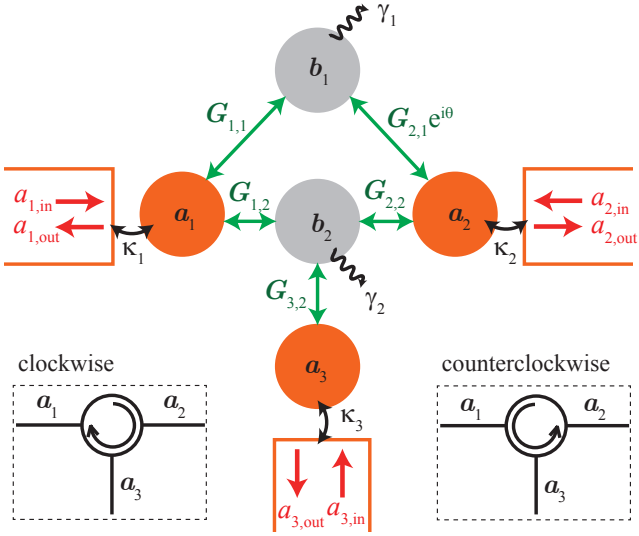}
\caption{(Color online) Schematic diagram of a three-port ($a_{1}$, $a_{2}$
and $a_{3}$) optical circulator by an electro-optomechanical system.}
\label{fig3}
\end{figure}

In the derivation of Sec.~III, we have assumed that $\kappa _{1}=\kappa
_{2}\ll \gamma _{2}$, where $\gamma _{2}$ should be the total damping rate
of the mechanical mode $2$. This assumption seems counterintuitive since usually the damping rate of the mechanical mode is smaller than the decay rate of the cavity mode. In this section, we will show that even when the intrinsic damping rate of the mechanical mode $2$ (denoted by $\gamma _{2,0}$)
is much smaller than the cavity decay rate $\kappa _{i}$, the total damping
rate of the mechanical mode $2$ can also satisfy the condition (\ref{Eq12})
when the mechanical resonator $2 $ is coupled to an auxiliary cavity mode
(cavity mode $3$), as shown in Fig.~\ref{fig3}. Moreover, we will
present the spectra of the output optical fields from the hybrid system
which involves the electro-optomechanical system and the auxiliary cavity
mode. We will show that the hybrid system can be used as a three-port
circulator for three optical modes with distinctively different wavelengths
at two different frequencies with opposite directions.

The Hamiltonian of the hybrid system for the electro-optomechanical system
with the auxiliary cavity mode is given by
\begin{equation}
H_{\mathrm{cir}}=H_{\mathrm{eom}}+H_{\mathrm{aux}},  \label{Eq39}
\end{equation}%
and
\begin{eqnarray}
H_{\mathrm{aux}} &=&\omega _{a,3}a_{3}^{\dag }a_{3}+g_{3,2}a_{3}^{\dag
}a_{3}\left( b_{2}+b_{2}^{\dag }\right)  \notag \\
&&+\Omega _{3,2}\left( a_{3}e^{i\left( \omega _{a,3}-\omega _{b,2}\right) t}+%
\mathrm{H.c.}\right),  \label{Eq40}
\end{eqnarray}%
where $a_{3}$ ($a_{3}^{\dag }$) is the bosonic annihilation (creation)
operator of the auxiliary cavity mode $3$ with resonance frequency $\omega _{a,3}$ and $g_{3,2}$ is the electromechanical (optomechanical) coupling strength between the cavity mode $3$ and the mechanical mode $2$.
The cavity mode $3$ is driven with strength $\Omega _{3,2}$ at frequency $%
\omega _{a,3}-\omega _{b,2}$. In the interaction picture with respect to $H_{%
\mathrm{cir,}0}=\sum_{i=1,2,3}\omega _{a,i}a_{i}^{\dag
}a_{i}+\sum_{j=1,2}\omega _{b,j}b_{j}^{\dag }b_{j}$, the linearized
Hamiltonian of Eq.~(\ref{Eq39}) can be written as%
\begin{equation}
H_{\mathrm{cir,int}}\approx H_{\mathrm{eom,int}}+G_{3,2}a_{3}^{\dag
}b_{2}+G_{3,2}a_{3}b_{2}^{\dag }  \label{Eq41}
\end{equation}
with the effective optomechanical coupling strength $G_{3,2}= g_{3,2}\alpha _{3,2}$. Without loss of generality, $G_{3,2}$ is assumed to be real.
The classical amplitude $\alpha _{3,2}$ is determined by solving the classical equation of motion with only the cavity drive $\Omega _{3,2}$ at frequency $\omega _{a,3}-\omega _{b,2}$.

The QLEs for the operators of the hybrid system is given as
\begin{equation}
\frac{d}{dt}V^{\prime \prime }\left( t\right) =-M^{\prime \prime }V^{\prime
\prime }\left( t\right) +\sqrt{\Gamma ^{\prime \prime }}V_{in}^{\prime
\prime }\left( t\right) ,  \label{Eq42}
\end{equation}%
with the vector $V^{\prime \prime }\left( t\right)
=\left( a_{1},a_{2},a_{3},b_{1},b_{2}\right) ^{T}$ of fluctuation operators, the
vector $V_{\mathrm{in}}^{\prime \prime }\left( t\right) =\left( a_{1,\mathrm{%
in}},a_{2,\mathrm{in}},a_{3,\mathrm{in}},b_{1,\mathrm{in}},b_{2,\mathrm{in}%
}\right) ^{T}$ of input operators, the diagonal damping matrix $\Gamma ^{\prime \prime }=%
\mathrm{diag}\left( \kappa _{1},\kappa _{2},\kappa _{3},\gamma _{1},\gamma
_{2,0}\right) $, and the coefficient matrix
\begin{equation}
M^{\prime \prime }=\left(
\begin{array}{ccccc}
\frac{\kappa _{1}}{2} & 0 & 0 & iG_{1,1} & iG_{1,2} \\
0 & \frac{\kappa _{2}}{2} & 0 & iG_{2,1}e^{i\theta } & iG_{2,2} \\
0 & 0 & \frac{\kappa _{3}}{2} & 0 & iG_{3,2} \\
iG_{1,1} & iG_{2,1}e^{-i\theta } & 0 & \frac{\gamma _{1}}{2} & 0 \\
iG_{1,2} & iG_{2,2} & iG_{3,2} & 0 & \frac{\gamma _{2,0}}{2}%
\end{array}%
\right) .  \label{Eq43}
\end{equation}%
Using the Fourier transform and the standard input-output relation, we can
express the output vector $V_{\mathrm{out}}^{\prime \prime }\left(t\right)
=\left( a_{1,\mathrm{out}},a_{2,\mathrm{out}},a_{3,\mathrm{out}},b_{1,%
\mathrm{out}},b_{2,\mathrm{out}}\right) ^{T}$ as
\begin{equation}
\widetilde{V}_{\mathrm{out}}^{\prime \prime }\left( \omega \right)
=U^{\prime \prime }\left( \omega \right) \widetilde{V}_{\mathrm{in}}^{\prime
\prime }\left( \omega \right) ,  \label{Eq44}
\end{equation}%
where%
\begin{equation}
U^{\prime \prime }\left( \omega \right) =\sqrt{\Gamma ^{\prime \prime }}%
\left( M^{\prime \prime }-i\omega I\right) ^{-1}\sqrt{\Gamma ^{\prime \prime
}}-I.  \label{Eq45}
\end{equation}

\begin{widetext}
\begin{figure*}[tbp]
\includegraphics[bb=30 328 551 576, width=16.5 cm, clip]{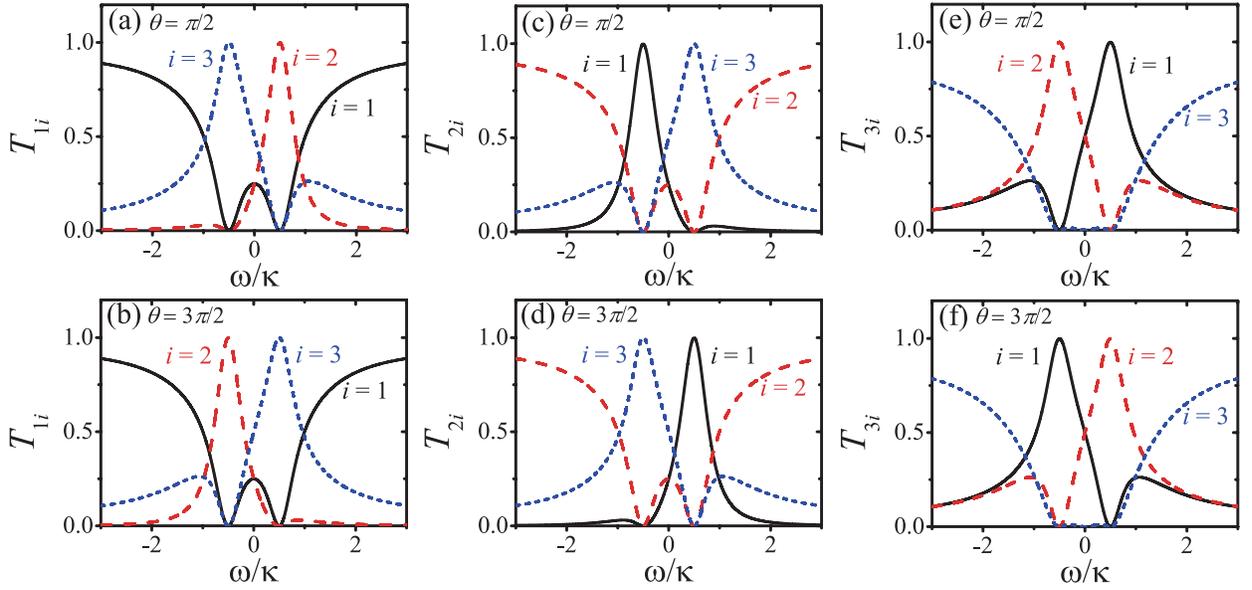}
\caption{(Color online) Scattering probabilities (a) and (b) $T_{1i}\left( \protect%
\omega \right)$, (c) and (d) $T_{2i}\left( \protect\omega \right)$, and (e) and (f) $T_{3i}\left(\protect\omega \right)$ ($i=1,2,3$)
as functions of the frequency of the incoming signal $\protect\omega$ for
different phase difference: (a), (c) and (e) $\protect\theta=\protect\pi/2$;
(b), (d) and (f) $\protect\theta =3\protect\pi/2$. The other parameters are $\kappa _{1}=\kappa _{2}=\kappa$, $\kappa _{3}=10\kappa$, $\gamma_{1}=\gamma_{2,0}=\kappa/1000$, $G_{1,1}=G_{2,1}=\kappa/2$, $G_{1,2}=G_{2,2}=2\kappa$, and $G_{3,2}=\sqrt{40}\kappa$ (thus, $\gamma _{2,{\rm id}}=16\kappa$).}
\label{fig4}
\end{figure*}
\end{widetext}

Under the assumption that the decay rate of the cavity mode $3$ is much
larger than the intrinsic damping rate of the mechanical mode $2$ and the
effective optomechanical coupling strength between the mechanical mode $2$
and the cavity mode $3$, i.e., $\kappa _{3}\gg \{\gamma _{2,0}, G_{3,2}\}$,
we can adiabatically eliminate the cavity mode $3$, then we obtained the
QLEs (\ref{Eq3}) with the replacement%
\begin{equation}
\gamma _{2}\rightarrow \gamma _{2,0}+\gamma _{2,\mathrm{id}}
\end{equation}%
in the coefficient matrix, and the replacement%
\begin{equation}
b_{2,\mathrm{in}}\rightarrow \sqrt{\gamma _{2,0}/\gamma _{2}}b_{2,\mathrm{in}%
}-i\sqrt{\gamma _{2,\mathrm{id}}/\gamma _{2}}a_{3,\mathrm{in}}  \label{Eq47}
\end{equation}%
in the input operators vector $V_{\mathrm{in}}\left( t\right) $. Here $%
\gamma _{2,\mathrm{id}}$ is the effective damping rate of the mechanical
mode $2$ induced by the auxiliary cavity mode $3$,
\begin{equation}
\gamma _{2,\mathrm{id}}=\frac{4G_{3,2}^{2}}{\kappa _{3}}.  \label{Eq48}
\end{equation}%
$\gamma _{2,\mathrm{id}}$ can be controlled by tuning the strength of the
driving field on the cavity mode $3$. Even if the intrinsic damping rate of
the mechanical mode $2$ is much smaller than the decay rates of the cavity
modes, i.e., $\gamma _{2,0}\ll \kappa _{i}$, the total damping rate of the
mechanical mode $2$ (i.e., $\gamma _{2}=\gamma _{2,0}+\gamma _{2,\mathrm{id}}
$) still can satisfy the condition (\ref{Eq12}) when $\gamma _{2,\mathrm{id}%
}\gg \kappa _{i}$.

In the following, we will study the scattering probability between the three
cavity modes. For convenience of discussion, we set $T_{ij}\left( \omega
\right) \equiv T_{a_{i},a_{j}}\left( \omega \right) =\left\vert
U_{ij}^{\prime \prime }\left( \omega \right) \right\vert ^{2}$ ($i,j=1,2,3$%
). Using Eq.~(\ref{Eq45}), we now show the numerical results of the
scattering probabilities between the three cavity modes. As shown in Fig.~%
\ref{fig4}, the electro-optomechanical system shows optical circulator
behavior for the three cavity modes at two different frequencies ($\omega
=\pm\kappa /2$) with opposite directions. When $\theta =\pi /2$ as shown in
Figs.~\ref{fig4} (a), (c) and (e), at frequency $\omega =-\kappa /2$, $%
T_{21}\left( \omega \right) \approx T_{32}\left( \omega \right) \approx
T_{13}\left( \omega \right) \approx 1$ and the other scattering
probabilities equal to zero; at frequency $\omega =\kappa /2$, $T_{12}\left(
\omega \right) \approx T_{23}\left( \omega \right) \approx T_{31}\left(
\omega \right) \approx 1$ and the other scattering probabilities equal to
zero. When $\theta =3\pi /2$, as shown in Figs.~\ref{fig4} (b), (d) and (f),
at frequency $\omega =-\kappa /2$, $T_{12}\left( \omega \right) \approx
T_{23}\left( \omega \right) \approx T_{31}\left( \omega \right) \approx 1$
and the other scattering probabilities equal to zero; at frequency $\omega
=\kappa /2$, $T_{21}\left( \omega \right) \approx T_{32}\left( \omega
\right) \approx T_{13}\left( \omega \right) \approx 1$ and the other
scattering probabilities equal to zero. That is when $\theta =\pi /2$, the
signal is transferred from one cavity mode to another either clockwisely ($%
a_{1}\rightarrow a_{2}\rightarrow a_{3}\rightarrow a_{1}$) at frequency $%
\omega =-\kappa /2$ or counterclockwisely ($a_{1}\rightarrow
a_{3}\rightarrow a_{2}\rightarrow a_{1} $) at frequency $\omega =\kappa /2$.
In contrast to $\theta =\pi /2$, when $\theta =3\pi /2$, the signal is
transferred either counterclockwisely at frequency $\omega =-\kappa /2$ or
clockwisely at frequency $\omega =\kappa /2$.

\section{Conclusions}

In summary, we have demonstrated the nonreciprocal conversion between microwave
and optical photons in electro-optomechanical systems. The
electro-optomechanical system shows nonreciprocal response between the
microwave and optical modes at two different frequencies with opposite
directions. The proposal is general and can be used to realize nonreciprocal
conversion between photons of two arbitrarily different frequencies.
Moreover, the electro-optomechanical system with an auxiliary optical mode
can be used as a three-port circulator for three optical modes with
arbitrarily different frequencies at two different frequencies with opposite directions.
The electro-optomechanical system with broken time-reversal symmetry will open up a different kind of quantum interface
in the quantum information processing and quantum networks.

\vskip 2pc \leftline{\bf Acknowledgement}

X.-W.X. thanks Professor Nian-Hua Liu for fruitful discussions. Y.L. is supported by the National Basic Research Program of China (973 Program) under
Grants No. 2014CB921403. A.-X.C. is supported by the National Natural Science
Foundation of China (NSFC) under Grants No. 11365009. Y.-x.L. is supported by NSFC under Grant Nos.
61328502 and 61025022, the Tsinghua University Initiative Scientific
Research Program, and the Tsinghua National Laboratory for Information
Science and Technology (TNList) Cross-discipline Foundation.
X.-W.X. is supported by the Startup Foundation for Doctors of East China Jiaotong University, under Grant No. 26541059.

\bibliographystyle{apsrev}
\bibliography{ref}

\begin{thebibliography}{99}
\bibitem{YouJQPT05} J. Q. You and F. Nori, \textit{Superconducting circuits and quantum information}, Phys. Today~\textbf{58}(11), 42 (2005).
\bibitem{DevoretSci13} M. H. Devoret and R. J. Schoelkopf, \textit{Superconducting Circuits for Quantum Information: An Outlook}, Science~\textbf{339}, 1169 (2013).

\bibitem{KimbleNat08} H. J. Kimble, \textit{The quantum internet}, Nature (London)~\textbf{453}, 1023 (2008).
\bibitem{RitterNat12} S. Ritter, C. Nolleke, C. Hahn, A. Reiserer, A. Neuzner, M. Uphoff, M. Mucke, E. Figueroa, J. Bochmann, and G. Rempe, \textit{An elementary quantum network of single atoms in optical cavities}, Nature (London)~\textbf{484}, 195 (2012).

\bibitem{XiangZLRMP13} Z. L. Xiang, S. Ashhab, J. Q. You, and F. Nori, \textit{Hybrid quantum circuits: Superconducting circuits interacting with other quantum systems}, Rev. Mod. Phys.~\textbf{85}, 623 (2013).
\bibitem{JZhouSR14} J. Zhou, Y. Hu, Z. Q. Yin, Z. D. Wang, S. L. Zhu, and Z. Y. Xue, \textit{High fidelity quantum state transfer in electromechanical systems with intermediate coupling}, Sci. Rep.~\textbf{4}, 6237 (2014).

\bibitem{KippenbergSci08} T. J. Kippenberg and K. J. Vahala, \textit{Cavity Optomechanics: Back-Action at the Mesoscale}, Science~\textbf{321}, 1172 (2008).
\bibitem{MarquardtPhy09} F. Marquardt and S. M. Girvin, \textit{Optomechanics}, Physics~\textbf{2}, 40 (2009).
\bibitem{AspelmeyerPT12} M. Aspelmeyer, P. Meystre, and K. Schwab, \textit{Quantum optomechanics}, Phys. Today~\textbf{65}(7), 29 (2012).
\bibitem{AspelmeyerARX13} M. Aspelmeyer, T. J. Kippenberg, and F. Marquardt, \textit{Cavity Optomechanics}, Rev. Mod. Phys.~\textbf{86}, 1391 (2014).



\bibitem{GroblacherNat09} S. Groblacher, K. Hammerer, M. R. Vanner, and M. Aspelmeyer, \textit{Observation of strong coupling between a micromechanical resonator and an optical cavity field}, Nature (London)~\textbf{460}, 724 (2009).
\bibitem{TeufelNat11a} J. D. Teufel, D. Li, M. S. Allman, K. Cicak, A. J. Sirois, J. D. Whittaker, and R. W. Simmonds, \textit{Circuit cavity electromechanics in the strong-coupling regime}, Nature (London)~\textbf{471}, 204 (2011).


\bibitem{TeufelNat11} J. D. Teufel, T. Donner, D. Li, J. W. Harlow, M. S. Allman, K. Cicak, A. J. Sirois, J. D. Whittaker, K. W. Lehnert, and R. W. Simmonds, \textit{Sideband cooling of micromechanical motion to the quantum ground state}, Nature (London)~\textbf{475}, 359 (2011).
\bibitem{ChanNat11} J. Chan, T. P. M. Alegre, A. H. Safavi-Naeini, J. T. Hill, A. Krause, S. Groblacher, M. Aspelmeyer, and O. Painter, \textit{Laser cooling of a nanomechanical oscillator into its quantum ground state}, Nature (London)~\textbf{478}, 89 (2011).
\bibitem{VerhagenNat12} E. Verhagen, S. Deleglise, S. Weis, A. Schliesser, and T. J. Kippenberg, \textit{Quantum-coherent coupling of a mechanical oscillator to an optical cavity mode}, Nature (London)~\textbf{482}, 63 (2012).



\bibitem{FiorePRL11} V. Fiore, Y. Yang, M. C. Kuzyk, R. Barbour, L. Tian, and H. Wang, \textit{Storing Optical Information as a Mechanical Excitation in a Silica Optomechanical Resonator}, Phys. Rev. Lett.~\textbf{107}, 133601 (2011).
\bibitem{PalomakiNat13} T. A. Palomaki, J. W. Harlow, J. D. Teufel, R. W. Simmonds, and K. W. Lehnert, \textit{Coherent state transfer between itinerant microwave fields and a mechanical oscillator}, Nature (London)~\textbf{495}, 210 (2013).


\bibitem{RegalJPCS11} C. A. Regal and K.W. Lehnert, \textit{From cavity electromechanics to cavity optomechanics},  J. Phys. Conf. Ser.~\textbf{264}, 012025 (2011).
\bibitem{TianAPB15} L. Tian, \textit{Optoelectromechanical transducer: Reversible conversion between microwave and optical photons}, Ann. Phys. (Berlin)~\textbf{527}, 1 (2015).


\bibitem{WangPRL12} Y. D. Wang and A. A. Clerk, \textit{Using Interference for High Fidelity Quantum State Transfer in Optomechanics}, Phys. Rev. Lett.~\textbf{108}, 153603 (2012).
\bibitem{TianPRL12} L. Tian, \textit{Adiabatic State Conversion and Pulse Transmission in Optomechanical Systems}, Phys. Rev. Lett.~\textbf{108}, 153604 (2012).
\bibitem{HKLiPRA13} H. K. Li, X. X. Ren, Y. C. Liu, and Y. F. Xiao, \textit{Photon-photon interactions in a largely detuned optomechanical cavity}, Phys. Rev. A~\textbf{88}, 053850 (2013).
\bibitem{ZhangKYPRL15} K. Y. Zhang, F. Bariani, Y. Dong, W. P. Zhang, and P. Meystre, \textit{Proposal for an Optomechanical Microwave Sensor at the Subphoton Level}, Phys. Rev. Lett.~\textbf{114}, 113601 (2015).

\bibitem{HillNC12} J. T. Hill, A. H. Safavi-Naeini, J. Chan, O. Painter, \textit{Coherent optical wavelength conversion via cavity optomechanics}, Nat. Comm.~\textbf{3}, 1196 (2012).
\bibitem{DongSci12} C. Dong, V. Fiore, M. C. Kuzyk, and H. Wang, \textit{Optomechanical Dark Mode}, Science~\textbf{338}, 1609 (2012).
\bibitem{LiuYPRL13} Y. Liu, M. Davan\c{c}o, V. Aksyuk, and K. Srinivasan, \textit{Electromagnetically Induced Transparency and Wideband Wavelength Conversion in Silicon Nitride Microdisk Optomechanical Resonators}, Phys. Rev. Lett.~\textbf{110}, 223603 (2013).

\bibitem{BagciNat14} T. Bagci, A. Simonsen, S. Schmid, L. G. Villanueva, E. Zeuthen, J. Appel, J. M. Taylor, A. S{\o }rensen, K.Usami, A. Schliesser, and E. S. Polzik, \textit{Optical detection of radio waves through a nanomechanical transducer}, Nature (London)~\textbf{507}, 81 (2014).

\bibitem{AndrewsNPy14} R. W. Andrews, R. W. Peterson, T. P. Purdy, K. Cicak, R. W. Simmonds, C. A. Regal, and K. W. Lehnert, \textit{Bidirectional and efficient conversion between microwave and optical light}, Nature Phys.~\textbf{10}, 321 (2014).


\bibitem{BochmannNPy13} J. Bochmann, A. Vainsencher, D. D. Awschalom, and A. N. Cleland, \textit{Nanomechanical coupling between microwave and optical photons},  Nature Phys.~\textbf{9}, 712 (2013);


\bibitem{BarzanjehPRL12} Sh. Barzanjeh, M. Abdi, G. J. Milburn, P. Tombesi, and D. Vitali, \textit{Reversible Optical-to-Microwave Quantum Interface}, Phys. Rev. Lett.~\textbf{109}, 130503 (2012)
\bibitem{WangYDPRL13} Y. D. Wang and A. A. Clerk, \textit{Reservoir-Engineered Entanglement in Optomechanical Systems}, Phys. Rev. Lett.~\textbf{110}, 253601 (2013).
\bibitem{LTianPRL13} L. Tian, \textit{Robust Photon Entanglement via Quantum Interference in Optomechanical Interfaces}, Phys. Rev. Lett.~\textbf{110}, 233602 (2013).
\bibitem{BarzanjehPRL15} Sh. Barzanjeh, S. Guha, C. Weedbrook, D. Vitali, J. H. Shapiro, and S. Pirandola, \textit{Microwave Quantum Illumination}, Phys. Rev. Lett.~\textbf{114}, 080503 (2015).


\bibitem{PottonRPP04} R. J. Potton, \textit{Reciprocity in optics}, Rep. Prog. Phys.~\textbf{67}, 717 (2004).
\bibitem{ShadrivovNJP11} I. V. Shadrivov, V. A. Fedotov, D. A. Powell, Y. S. Kivshar, and N. I. Zheludev, \textit{Electromagnetic wave analogue of an electronic diode}, New J. Phys.~\textbf{13}, 033025 (2011).


\bibitem{FujitaAPL00} J. Fujita, M. Levy, R. M. Osgood, L.Wilkens, and H. D\"{o}tsch, \textit{Waveguide optical isolator based on Mach-Zehnder interferometer}, Appl. Phys. Lett.~\textbf{76}, 2158 (2000).
\bibitem{EspinolaOL04} R. L. Espinola, T. Izuhara, M. C. Tsai, R. M. Osgood Jr., H. D\"{o}tsch, \textit{Magneto-optical nonreciprocal phase shift in garnet/silicon-on-insulator waveguides}, Opt. Lett.~\textbf{29}, 941 (2004).
\bibitem{ZamanAPL07} T. R. Zaman, X. Guo, R. J. Ram, \textit{Faraday rotation in an InP waveguide}, Appl. Phys. Lett.~\textbf{90}, 023514 (2007).
\bibitem{HaldanePRL08} F. D. M. Haldane and S. Raghu, \textit{Possible Realization of Directional Optical Waveguides in Photonic Crystals with Broken Time-Reversal Symmetry}, Phys. Rev. Lett.~\textbf{100}, 013904 (2008).
\bibitem{ShojiAPL08} Y. Shoji, T. Mizumoto, H. Yokoi, I. Hsieh, and R. M. Osgood Jr., \textit{Magneto-optical isolator with silicon waveguides fabricated by direct bonding}, Appl. Phys. Lett.~\textbf{92}, 071117 (2008).
\bibitem{ZWangNat09} Z. Wang, Y. Chong, J. D. Joannopoulos, and M. Solja\v{c}i\'{c}, \textit{Observation of unidirectional backscattering-immune topological electromagnetic states}, Nature (London)~\textbf{461}, 772 (2009).
\bibitem{HadadPRL10} Y. Hadad and B. Z. Steinberg, \textit{Magnetized Spiral Chains of Plasmonic Ellipsoids for One-Way Optical Waveguides}, Phys. Rev. Lett.~\textbf{105}, 233904 (2010).
\bibitem{KhanikaevPRL10} A. B. Khanikaev, S. H. Mousavi, G. Shvets, and Y. S. Kivshar, \textit{One-Way Extraordinary Optical Transmission and Nonreciprocal Spoof Plasmons}, Phys. Rev. Lett.~\textbf{105}, 126804 (2010).
\bibitem{LBiNPo11} L. Bi, J. Hu, P. Jiang, D. H. Kim, G. F. Dionne, L. C. Kimerling, and C. A. Ross, \textit{On-chip optical isolation in monolithically integrated non-reciprocal optical resonators}, Nat. Photon.~\textbf{5}, 758 (2011).
\bibitem{ShojiOE12} Y. Shoji, M. Ito, Y. Shirato, and T. Mizumoto, \textit{MZI optical isolator with Si-wire waveguides by surface-activated direct bonding}, Opt. Express~\textbf{20}, 18440 (2012).


\bibitem{GalloAPL01} K. Gallo, G. Assanto, K. R. Parameswaran, and M. M. Fejer, \textit{All-optical diode in a periodically poled lithium niobate waveguide}, Appl. Phys. Lett.~\textbf{79}, 314 (2001).
\bibitem{MingaleevJOSAB02} S. F. Mingaleev, Y. S. Kivshar, \textit{Nonlinear transmission and light localization in photonic-crystal waveguides}, J. Opt. Soc. Am. B~\textbf{19}, 2241 (2002).
\bibitem{SoljacicOL03} M. Solja\v{c}i\'{c}, C. Luo, J. D. Joannopoulos, S. Fan, \textit{Nonlinear photonic crystal microdevices for optical integration}, Opt. Lett.~\textbf{28}, 637 (2003).
\bibitem{RostamiOLT07} A. Rostami, \textit{Piecewise linear integrated optical device as an optical isolator using two-port nonlinear ring resonators}, Opt. Laser Technol.~\textbf{39}, 1059 (2007).
\bibitem{AlberucciOL08} A. Alberucci and G. Assanto, \textit{All-optical isolation by directional coupling}, Opt. Lett.~\textbf{33}, 1641 (2008).
\bibitem{LFanSci12} L. Fan, J. Wang, L. T. Varghese, H. Shen, B. Niu, Y. Xuan, A. M. Weiner, and M. Qi, \textit{An All-Silicon Passive Optical Diode}, Science~\textbf{335}, 447 (2012).
\bibitem{LFanOL13} L. Fan, L. T. Varghese, J. Wang, Y. Xuan, A. M. Weiner, and M. Qi, \textit{Silicon optical diode with 40 dB nonreciprocal transmission}, Opt. Lett.~\textbf{38}, 1259 (2013).
\bibitem{AnandNL13} B. Anand, R. Podila, K. Lingam, S. R. Krishnan, S. S. S. Sai, R. Philip, and A. M. Rao, \textit{Optical Diode Action from Axially Asymmetric Nonlinearity in an All-Carbon Solid-State Device}, Nano Lett.~\textbf{13}, 5771 (2013).


\bibitem{BiancalanaJAP08} F. Biancalana, \textit{All-optical diode action with quasiperiodic photonic crystals}, J. Appl. Phys.~\textbf{104}, 093113 (2008).
\bibitem{MiroshnichenkoAPL10} A. E. Miroshnichenko, E. Brasselet, and Y. S. Kivshar, \textit{Reversible optical nonreciprocity in periodic structures with liquid crystals}, Appl. Phys. Lett.~\textbf{96}, 063302 (2010).
\bibitem{CWangOE11} C. Wang, C. Zhou, and Z. Li, \textit{On-chip optical diode based on silicon photonic crystal heterojunctions}, Opt. Express~\textbf{19}, 26948 (2011).
\bibitem{CWangSR12} C. Wang, X. Zhong, and Z. Li, \textit{Linear and passive silicon optical isolator}, Sci. Rep.~\textbf{2}, 674 (2012).
\bibitem{KXiaOE13} K. Xia, M. Alamri, and M. S. Zubairy, \textit{Ultrabroadband nonreciprocal transverse energy flow of light in linear passive photonic circuits}, Opt. Express~\textbf{21}, 25619 (2013).
\bibitem{LenferinkOE14} E. J. Lenferink, G. Wei, and N. P. Stern, \textit{Coherent optical non-reciprocity in axisymmetric resonators}, Opt. Express~\textbf{22}, 16099 (2014).
\bibitem{YYuarX14} Y. Yu, Y. Chen, H. Hu, W. Xue, K. Yvind, and J. Mork, \textit{Nonreciprocal transmission in a nonlinear photonic-crystal Fano structure with broken symmetry}, Laser Photonics Rev.~\textbf{9}, 241 (2015).


\bibitem{YuNP09} Z. F. Yu and S. H. Fan, \textit{Complete optical isolation created by indirect interband photonic transitions}, Nat. Photon.~\textbf{3}, 91 (2009).
\bibitem{KFangNPo12} K. Fang, Z. Yu, and S. Fan, \textit{Realizing effective magnetic field for photons by controlling the phase of dynamic modulation}, Nat. Photon.~\textbf{6}, 782 (2012).
\bibitem{ELiNC14} E. Li, B. J. Eggleton, K. Fang, and S. Fan, \textit{Photonic Aharonov-Bohm effect in photon-phonon interactions}, Nat. Commun.~\textbf{5}, 3225 (2014).
\bibitem{DoerrOL11} C. R. Doerr, N. Dupuis, and L. Zhang, \textit{Optical isolator using two tandem phase modulators}, Opt. Lett.~\textbf{36}, 4293 (2011).
\bibitem{DoerrOE14} C. R. Doerr, L. Chen, and D. Vermeulen, \textit{Silicon photonics broadband modulation-based isolator}, Opt. Express~\textbf{22}, 4493 (2014).
\bibitem{LiraPRL12} H. Lira, Z. F. Yu, S. H. Fan, and M. Lipson, \textit{Electrically Driven Nonreciprocity Induced by Interband Photonic Transition on a Silicon Chip}, Phys. Rev. Lett.~\textbf{109}, 033901 (2012).
\bibitem{TzuangNPt14} L. D. Tzuang, K. Fang, P. Nussenzveig, S. Fan, and M. Lipson, \textit{Non-reciprocal phase shift induced by an effective magnetic flux for light}, Nat. Photon.~\textbf{8}, 701 (2014).
\bibitem{MunozPRL14} M. Castellanos Munoz, A. Y. Petrov, L. O'Faolain, J. Li, T. F. Krauss, and M. Eich, \textit{Optically Induced Indirect Photonic Transitions in a Slow Light Photonic Crystal Waveguide}, Phys. Rev. Lett.~\textbf{112}, 053904 (2014).
\bibitem{YYangOE14} Y. Yang, C. Galland, Y. Liu, K. Tan, R. Ding, Q. Li, K. Bergman, T. Baehr-Jones, and M. Hochberg, \textit{Experimental demonstration of broadband Lorentz non-reciprocity in an integrable photonic architecture based on Mach-Zehnder modulators}, Opt. Express~\textbf{22}, 17409 (2014).
\bibitem{MSKangNP11} M. S. Kang, A. Butsch, and P. S. J. Russell, \textit{Reconfigurable light-driven opto-acoustic isolators in photonic crystal fibre}, Nat. Photonics~\textbf{5}, 549 (2011).
\bibitem{EuterNP10} C. E\"{u}ter, K. G. Makris, R. EI-Ganainy, D. N. Christodoulides, M. Segev, and D. Kip, \textit{Observation of parity-time symmetry in optics}, Nat. Phys.~\textbf{6}, 192 (2010).
\bibitem{LFengSci11} L. Feng, M. Ayache, J. Q. Huang, Y. L. Xu, M. H. Lu, Y. F. Chen, Y. Fainman, and A. Scherer, \textit{Nonreciprocal Light Propagation in a Silicon Photonic Circuit}, Science~\textbf{333}, 729 (2011).
\bibitem{BPengNP14} B. Peng, S. K. \"{O}zdemir, F. Lei, F. Monifi, M. Gianfreda, G. L. Long, S. H. Fan, F. Nori, C. M. Bender, and L. Yang, \textit{Parity-time-symmetric whispering-gallery microcavities}, Nat. Phys.~\textbf{10}, 394 (2014).
\bibitem{WangPRL13} D. W. Wang, H. T. Zhou, M. J. Guo, J. X. Zhang, J. Evers, and S. Y. Zhu, \textit{Optical Diode Made from a Moving Photonic Crystal}, Phys. Rev. Lett.~\textbf{110}, 093901 (2013).

\bibitem{WangOE10} Q. Wang, F. Xu, Z. Y. Yu, X. S. Qian, X. K. Hu, Y. Q. Lu, and H. T. Wang, \textit{A bidirectional tunable optical diode based on periodically poled LiNbO3}, Opt. Express~\textbf{18}, 7340 (2010).
\bibitem{RamezaniPRA10} H. Ramezani, T. Kottos, R. El-Ganainy, and D. N. Christodoulides, \textit{Unidirectional nonlinear $\mathcal{PT} $-symmetric optical structures}, Phys. Rev. A~\textbf{82}, 043803 (2010).
\bibitem{KFangPRL12} K. Fang, Z. Yu, and S. Fan, \textit{Photonic Aharonov-Bohm Effect Based on Dynamic Modulation}, Phys. Rev. Lett.~\textbf{108}, 153901 (2012).
\bibitem{HorsleyPRL13} S. A. R. Horsley, J. H. Wu, M. Artoni, and G. C. La Rocca, \textit{Optical Nonreciprocity of Cold Atom Bragg Mirrors in Motion}, Phys. Rev. Lett.~\textbf{110}, 223602 (2013).
\bibitem{JHWuPRL14} J. H. Wu, M. Artoni, and G. C. La Rocca, \textit{Non-Hermitian Degeneracies and Unidirectional Reflectionless Atomic Lattices}, Phys. Rev. Lett.~\textbf{113}, 123004 (2014).


\bibitem{ManipatruniPRL09} S. Manipatruni, J. T. Robinson, and M. Lipson, \textit{Optical Nonreciprocity in Optomechanical Structures}, Phys. Rev. Lett.~\textbf{102}, 213903 (2009).

\bibitem{HafeziOE12} M. Hafezi and P. Rabl, \textit{Optomechanically induced non-reciprocity in microring resonators}, Opt. Express~\textbf{20}, 7672 (2012).

\bibitem{KimNPy15} J. Kim, M. C. Kuzyk, K. Han, H. Wang, and G. Bahl, \textit{Non-reciprocal Brillouin scattering induced transparency}, Nat. Phys.~\textbf{11}, 275 (2015).
\bibitem{CHDongNC15} C. H. Dong, Z. Shen, C. L. Zou, Y. L. Zhang, W. Fu, and G. C. Guo, \textit{Brillouin-scattering-induced transparency and non-reciprocal light storage}, Nature Commun.~\textbf{6}, 6193 (2015).


\bibitem{XuXWPRA15} X. W. Xu and Y. Li, \textit{Optical nonreciprocity and optomechanical circulator in three-mode optomechanical systems}, Phys. Rev. A~\textbf{91}, 053854 (2015).

\bibitem{KochPRA10} J. Koch, A. A. Houck, K. L. Hur, and S. M. Girvin, \textit{Time-reversal-symmetry breaking in circuit-QED-based photon lattices}, Phys. Rev. A~\textbf{82}, 043811 (2010).
\bibitem{HabrakenNJP12} S. J. M. Habraken, K. Stannigel, M. D. Lukin, P. Zoller, and P Rabl, \textit{Continuous mode cooling and phonon routers for phononic quantum networks}, New J. Phys.~\textbf{14}, 115004 (2012).
\bibitem{SliwaPRX15} K. M. Sliwa, M. Hatridge, A. Narla, S. Shankar, L. Frunzio, R. J. Schoelkopf, and M. H. Devoret, \textit{Reconfigurable Josephson Circulator/Directional Amplifier}, Phys. Rev. X~\textbf{5}, 041020 (2015).
\bibitem{SchmidtOpt15} M. Schmidt, S. Kessler, V. Peano, O. Painter, and F. Marquardt, \textit{Optomechanical creation of magnetic fields for photons on a lattice}, Optica~\textbf{2}, 635 (2015).

\bibitem{FangKArx15} K. Fang, M. H. Matheny, X. Luan, and O. Painter, \textit{Phonon routing in integrated optomechanical cavity-waveguide systems}, arXiv:1508.05138v1 [physics.optics].


\bibitem{MetelmannPRX15} A. Metelmann and A. A. Clerk, \textit{Nonreciprocal Photon Transmission and Amplification via Reservoir Engineering}, Phys. Rev. X~\textbf{5}, 021025 (2015).


\bibitem{KronwaldPRA13} A. Kronwald, F. Marquardt and A. A. Clerk, \textit{Arbitrarily large steady-state bosonic squeezing via dissipation}, Phys. Rev. A~\textbf{88}, 063833 (2013).
\bibitem{OjanenPRA14} T. Ojanen and K. B{\o}rkje, \textit{Ground-state cooling of mechanical motion in the unresolved sideband regime by use of optomechanically induced transparency}, Phys. Rev. A~\textbf{90}, 013824 (2014).

\bibitem{DeJesusPRA87} E. X. DeJesus and C. Kaufman, \textit{Routh-Hurwitz criterion in the examination of eigenvalues of a system of nonlinear ordinary differential equations}, Phys. Rev. A~\textbf{35}, 5288 (1987).
\bibitem{Gradshteyn80} I. S. Gradshteyn and I. M. Ryzhik, in \textit{Table of Integrals, Series and Products} (Academic, Orlando, 1980), p. 1119.
\bibitem{PaternostroNJP06} M. Paternostro, S. Gigan, M. S. Kim, F. Blaser, H. R. B\"{o}hm, and M. Aspelmeyer, \textit{Reconstructing the dynamics of a movable mirror in a detuned optical cavity}, New J. Phys.~\textbf{8}, 107 (2006).
\bibitem{VitaliPRL07} D. Vitali, S. Gigan, A. Ferreira, H. R. B\"{o}hm, P. Tombesi, A. Guerreiro, V. Vedral, A. Zeilinger, and M. Aspelmeyer, \textit{Optomechanical Entanglement between a Movable Mirror and a Cavity Field}, Phys. Rev. Lett.~\textbf{98}, 030405 (2007).
\bibitem{GhobadiPRA11} R. Ghobadi, A. R. Bahrampour, and C. Simon, \textit{Quantum optomechanics in the bistable regime}, Phys. Rev. A~\textbf{84}, 033846 (2011).

\bibitem{GardinerPRA85} C. W. Gardiner and M. J. Collett, \textit{Input and output in damped quantum systems: Quantum stochastic differential equations and the master equation}, Phys. Rev. A~\textbf{31}, 3761 (1985).

\bibitem{AgarwalPRA12} G. S. Agarwal and S. Huang, \textit{Optomechanical systems as single-photon routers}, Phys. Rev. A~\textbf{85}, 021801(R) (2012). %

\bibitem{JahnePRA09} K. J\"{a}hne, C. Genes, K. Hammerer, M. Wallquist, E. S. Polzik, and P. Zoller, \textit{Cavity-assisted squeezing of a mechanical oscillator}, Phys. Rev. A~\textbf{79}, 063819 (2009).
\bibitem{XWXuPRA15} X. W. Xu, Y. X. Liu, C. P. Sun, and Y. Li, \textit{Mechanical $\mathcal{PT}$ symmetry in coupled optomechanical systems}, Phys. Rev. A~\textbf{92}, 013852 (2015).

\end{thebibliography}

\end{document}